\documentclass[journal]{IEEEtran}

\usepackage{cite}
\usepackage{amsmath,amssymb,amsfonts}
\usepackage{algorithmic}
\usepackage{verbatim}
\usepackage{graphicx}
\usepackage{textcomp}
\usepackage{xcolor}
\usepackage{kotex}
\usepackage{ctable}
\usepackage{multirow}
\usepackage{multicol}
\usepackage{tabularx}
\usepackage{array}
\usepackage{xspace}
\usepackage{mathtools}
\usepackage[normalem]{ulem}
\usepackage[hyphens]{url}
\usepackage{datetime}
\usepackage{fancyhdr}
\usepackage[caption=false,labelfont=sf,textfont=sf]{subfig}
\usepackage{enumitem}
\usepackage{stfloats}
\usepackage{makecell}
\usepackage{listings}

\definecolor{codebg}{rgb}{0.97,0.97,0.97}

\newcolumntype{Y}{>{\centering\arraybackslash}X}

\newcommand{\sys}{\textsf{\footnotesize IoTGPT}\xspace}
\newcommand{\syscold}{\textsf{\footnotesize IoTGPT-Cold}\xspace}

\newcommand{\syswarm}{\textsf{\footnotesize IoTGPT-Warm}\xspace}
\newcommand{\sasha}{\textsf{\footnotesize Sasha}\xspace}
\newcommand{\sage}{\textsf{\footnotesize SAGE}\xspace}
\newcommand{\nodecomp}{\textsf{\footnotesize NoDecomp}\xspace}
\newcommand{\nomem}{\textsf{\footnotesize NoMem}\xspace}
\newcommand{\baseenv}{\textsf{\footnotesize BASE}\xspace}
\newcommand{\newenv}{\textsf{\footnotesize NEW}\xspace}

\newcommand{\newitem}[1]{\vspace{1mm} \noindent\textbf{#1}}
\newenvironment{sitemize}{
  \begin{list}{$\bullet$}{
    \setlength{\leftmargin}{1em}}%
  }{\end{list}}

\begin{document}

\title{Leveraging LLMs for Efficient and Personalized Smart Home Automation}

\author{
Chaerin~Yu,
Chihun~Choi,
Sunjae Lee,
Hyosu~Kim,
Steven~Y.~Ko,
Young-Bae~Ko,
and~Sangeun~Oh
\thanks{C. Yu and Y. Ko are with the Department of AI Convergence Network, Ajou University, Suwon 16499, South Korea (e-mail: dbcofls6961@ajou.ac.kr; youngko@ajou.ac.kr).}
\thanks{S. Lee is with the Department of Computer Science and Engineering, Sungkyunkwan University, Suwon 16419, South Korea (e-mail: sunjae.lee@skku.edu).}
\thanks{H. Kim is with the School of Computer Science and Engineering, Chung-Ang University, Seoul 06974, South Korea (e-mail: hskimhello@cau.ac.kr).}
\thanks{S. Y. Ko is with the School of Computing Science, Simon Fraser University, Burnaby V5A 1S6, BC, Canada (e-mail: steveyko@sfu.ca).}
\thanks{C. Choi and S. Oh are with the Department of Computer Science and Engineering, Korea University, Seoul 02841, South Korea (e-mail: clgns0102@korea.ac.kr ; sangeunoh@korea.ac.kr).}
\thanks{Sangeun Oh and Young-Bae Ko are co-corresponding authors.}
}



\maketitle
\begin{abstract}
The proliferation of smart home devices has increased the complexity of controlling and managing them, leading to user fatigue. In this context, large language models (LLMs) offer a promising solution by enabling natural-language interfaces for Internet of Things (IoT) control. However, existing LLM-based approaches suffer from unreliable and inefficient device control due to the non-deterministic nature of LLMs, high inference latency and cost, and limited personalization. To address these challenges, we present \sys{}, an LLM-based smart home agent designed to execute IoT commands in a reliable, efficient, and personalized manner. Inspired by how humans manage complex tasks, \sys{} decomposes user instructions into subtasks and memorizes them. By reusing learned subtasks, subsequent instructions can be processed more efficiently with fewer LLM calls, improving reliability and reducing both latency and cost. \sys{} also supports fine-grained personalization by adapting individual subtasks to user preferences. Our evaluation demonstrates that \sys{} outperforms baselines in accuracy, latency/cost, and personalization, while reducing user workload.

\end{abstract}

\section{Introduction}

\IEEEPARstart{O}{ver} the past decade, the growth of IoT technologies has brought a variety of devices---ranging from lighting and heating to security systems and home appliances---into smart homes, making these environments increasingly complex. As of 2025, households in the United States owned an average of 21 connected IoT devices~\cite{survey2025}, underscoring the scale of this complexity. With this growing number of interconnected devices, users face a heightened cognitive load in controlling and managing them effectively~\cite{bainbridge1982ironies}. Consequently, there is an increasing demand for efficient and intuitive approaches to IoT device control that can help alleviate user fatigue.

Unsurprisingly, various methods have been explored to facilitate smart home automation. Existing IoT platforms such as Samsung SmartThings~\cite{SmartThings}, Google Home~\cite{GoogleHome}, and Apple HomeKit~\cite{AppleHome} provide visual interfaces to define trigger-action rules without writing source code. While such interfaces are intuitive for handling simple tasks, they impose increasing cognitive load as the desired automation grows in complexity. Alternatively, voice assistants, such as Google Assistant~\cite{GoogleAssistant} and Apple Siri~\cite{Siri}, can translate users' spoken instructions into IoT commands. These systems perform well when interpreting explicit and well-structured requests (e.g., ``\textit{Turn on the living room lights at 6 p.m.}'') based on predefined instruction templates. However, they struggle with more natural but ambiguous instructions (e.g., ``\textit{Make the living room bright after sunset}'') where reasoning about user intent is required, leading to incorrect or incomplete automation.

Large language models (LLMs) are emerging as a game-changer for smart home automation. Unlike traditional approaches, LLM-based agents~\cite{Sasha24, SAGE25, AwareAuto24, RuleBot++24, Calo2024PersonalUbiqCom} can flexibly interpret user utterances and reason about underlying intent, deriving IoT commands even from ambiguous instructions. However, they also face key limitations. First, the non-deterministic nature of LLMs undermines reliability: hallucinated commands require users to repeatedly review and correct results. Second, LLM inference is computationally expensive, with long prompt processing and reasoning. Existing agents exacerbate this cost by repeatedly invoking the LLM through multi-step reasoning and replanning; in our experiments, one approach took up to 151 seconds to process a simple instruction. Lastly, LLMs lack user-specific context, making it difficult to personalize automation. For example, for an underspecified instruction like ``\textit{Make the bedroom ready for sleep},'' users often have different preferences about what this entails (e.g., dimmer lighting, cooler temperature, or lower background noise), which current systems struggle to infer reliably.

To overcome these limitations, we introduce \sys{}, an LLM-based smart home agent that translates natural language instructions into executable IoT commands in a reliable, efficient, and personalized manner. The key idea is \textit{task decomposition and memorization}. Inspired by how humans break down complex tasks into smaller, manageable subtasks and learn them for future use~\cite{correa2023humans, lovden2020human}, \sys{} decomposes an instruction into explicit subtasks (e.g., ``\textit{make the bedroom ready for sleep}'' $\rightarrow$ \{``\textit{adjust air conditioner temperature},'' ``\textit{dim the sleep light}''\}) that can be easily translated into API-level commands. These subtasks are then stored in memory, along with their translated device commands, and reused when similar device control subtasks appear in future instructions.

This decomposition enables two major benefits: \textit{i)} subtasks can be memorized and reused across different instructions, reducing repeated LLM inference and thereby mitigating unreliability, latency, and cost in command generation; and \textit{ii)} configurable parameters exposed through subtasks allow low-level device commands to be fine-tuned based on given context and user preferences. Moreover, preferences learned in one subtask can guide others. For instance, if \sys learns that a user's preference for ``cool'' corresponds to 22°C while controlling an air conditioner, it can automatically apply this same preference when adjusting room temperature using other cooling devices such as a fan, enabling adaptive personalization even when device configurations change.

To achieve our goal, \sys{} introduces three key features: \textit{i) Reliable task decomposition:} Due to their stochastic nature, LLMs may generate inconsistent subtasks or invalid device commands. To mitigate this, \sys{} integrates structured prompting with correction mechanisms to ensure valid commands. \textit{ii) Efficient subtask reuse:} To facilitate efficient subtask reuse, \sys{} organizes task knowledge into an abstracted, hierarchical data structure (\textit{task memory}). Task memory compactly stores subtasks and device commands in generalized form, enabling flexible reuse across tasks. \textit{iii) Adaptive personalization:} To enable personalization even when device availability differs, \sys{} abstracts user preferences from device-specific settings into device-agnostic environmental properties. This allows preferences to transfer across devices; for example, if a user's preference for a cool environment is learned from air conditioner usage, \sys{} can approximate it through fan speed adjustment when only a fan is available.

\newitem{Contributions.} In summary, our contributions are as follows:
\begin{sitemize}
    \item We propose \sys{}, an LLM-based smart home agent that leverages task decomposition and memorization for reliable, efficient, and personalized IoT control.
    
    \item We introduce a hierarchical \textit{task memory} that enables efficient subtask reuse, reducing reliance on the LLM and improving both reliability and efficiency in automation.
    
    \item We design adaptive personalization that abstracts preferences into device-agnostic environmental properties, enabling transfer across heterogeneous device configurations.
    
    \item We conducted experiments with a prototype of \sys{} on the Samsung SmartThings platform, showing up to 85.43\% higher successful task rates, 78.40\% lower latency and 44\% lower cost compared to state-of-the-art baselines.

    \item We performed in-depth user studies with 15 participants to evaluate \sys{}. The results show reduced cognitive load, improved usability, and more effective personalization compared to existing approaches.

\end{sitemize}

\section{Related Work} \label{sec:related work}
\sys{} is an LLM-based agent for smart home automation. Given this scope, we review related work in three areas: \textit{traditional smart home automation}, \textit{LLM-based smart home agents}, and \textit{LLM-based task automation}.

\newitem{Traditional smart home automation.}
Automation has become a rapidly growing demand in smart homes, driving the development of diverse automation techniques~\cite{SH-RDL25}. Traditional platforms (e.g., SmartThings~\cite{SmartThings}, Google Home~\cite{GoogleHome}, Apple HomeKit~\cite{AppleHome}) require users to manually create trigger-action rules through visual interfaces. While simple, these rules limit flexibility and impose high cognitive load as task complexity increases. To ease this burden, crowdsourcing-based solutions~\cite{InstructableCrowd16, Almond17} share rules across users, yet suffer from reliability issues due to quality control and standardization challenges. Machine learning approaches~\cite{Jaihar20} attempt to personalize control by inferring user routines from sensor and device logs, yet often fail to capture user intent. Natural language interfaces~\cite{COMMA22, RuleBot16, HeyTAP21} have also been integrated into assistants like Siri~\cite{Siri} and Google Assistant~\cite{GoogleAssistant}, providing intuitive control through spoken instructions. However, their reliance on predefined instruction templates limits adaptation to informal expressions, which are often ambiguous, thereby reducing usability. Consequently, traditional approaches remain rigid and burdensome, struggling with complex or ambiguous requests. This underscores the need for systems that can autonomously handle intent interpretation and adaptation, minimizing reliance on manual user intervention.

\newitem{LLM-based smart home agents.} 
Recent studies have explored leveraging LLMs for smart home automation~\cite{AwareAuto24, RuleBot++24, Sasha24, SAGE25, Calo2024PersonalUbiqCom}. 
Some works~\cite{RuleBot++24, AwareAuto24} translate diverse natural language requests into automation rules via LLMs' reasoning capabilities, while \cite{Calo2024PersonalUbiqCom} leverages LLMs to present visual options for instruction disambiguation. However, these approaches entail significant user involvement, requiring users to either validate generated rules or manually select the intended interpretation. Sasha~\cite{Sasha24} reduces this burden through multi-step reasoning such as user request clarification and relevant device filtering. However, it still lacks consideration of user context and personalization. Moreover, its monolithic approach, processing an entire instruction in a single reasoning pass, makes it prone to failures particularly for complex tasks. To address this, SAGE~\cite{SAGE25} dynamically plans sequences of reasoning modules to flexibly handle tasks and leverages past device interactions for personalization. However, this heavy reliance on the LLM introduces both computational overhead and error accumulation due to its non-deterministic nature. Furthermore, its device-specific personalization limits adaptability when device configurations change.
In contrast, \sys{} minimizes repeated LLM reasoning by decomposing instructions into reusable subtasks and caching them for future use. Furthermore, unlike prior works limited to device-specific history, \sys{} abstracts user preferences into device-agnostic properties, enabling flexible personalization across heterogeneous devices.

\newitem{LLM-based task automation.}
LLMs have been explored as a solution for end-to-end task automation in diverse computing domains. For instance, mobile agents~\cite{MobileGPT24, AutoDroid, AppAgent, lee2025safeguarding, wen2024autodroid} automate the control of mobile apps by leveraging LLMs to interpret GUI components. Another line of work~\cite{nakano2021webgpt, WebArena, hong2024cogagent, tomar2025cybernaut} focuses on task automation in web browsers through the analysis of web components, such as HTML DOM tree structures. Similarly, LLMs have automated operations within desktop productivity software by monitoring a user's screen~\cite{ASSISTGUI, niu2024screenagent, zhang2025ufo2, zhao2025cola}. These studies demonstrate LLMs' potential for task automation in software environments through screen interpretation. Notably, similar to \sys{}, some of these works adopt task decomposition strategies to tackle complex workflows. However, smart home automation targeted by \sys{} introduces distinct challenges: it requires reasoning over dynamic physical contexts (e.g., environmental conditions), coordinating heterogeneous devices, and supporting adaptive personalization. \sys{} tackles these unique complexities by integrating task decomposition with context-aware memory, ensuring reliable and personalized automation in physical spaces.

\section{IoTGPT Design}

We present \sys{}, an LLM-based smart home agent designed for reliable, efficient, and personalized IoT control. This section details its design principles and challenges.

\begin{figure*}[t]               
  \centering
  \includegraphics[width=0.9\linewidth]{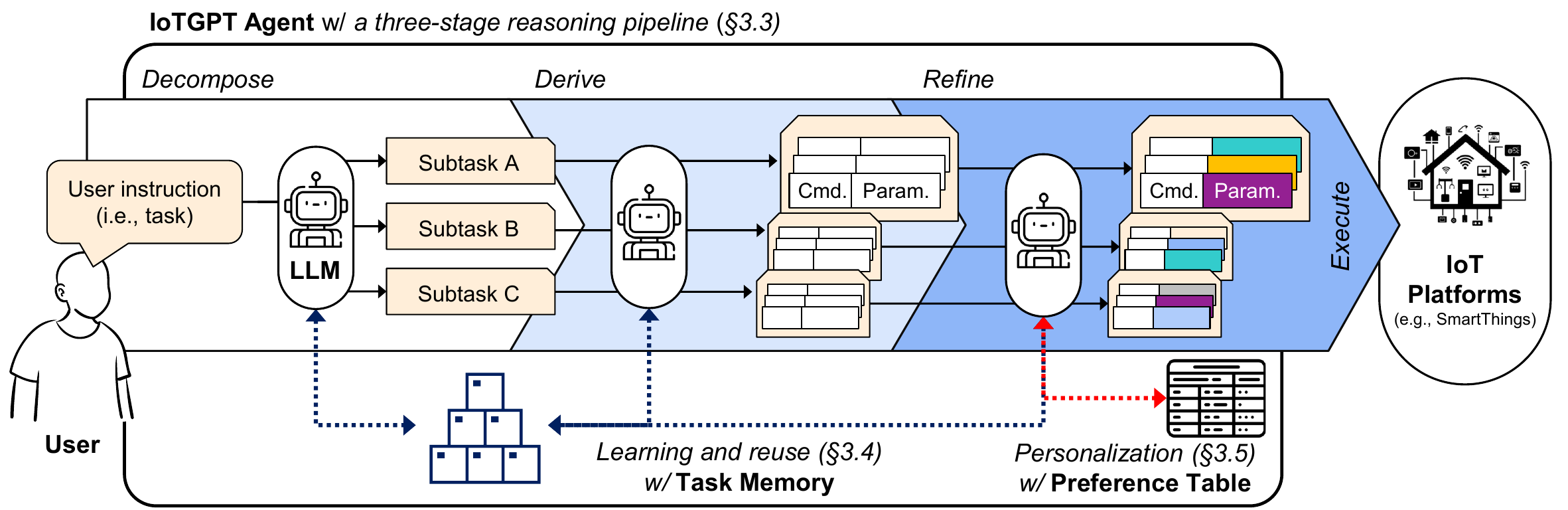}
  \caption{System overview of \sys{}, which translates natural-language instructions into IoT commands via a three-stage reasoning pipeline (Decompose–Derive–Refine), leveraging task memory for learning/reuse and a preference table for personalization.
  }
  \label{fig:architecture}
  \vspace{-0.3cm}
\end{figure*}

\subsection{System Design Overview}

\newitem{Design principles.}
When faced with a complex and ambiguous task, people rarely solve it in one shot; instead, they decompose it into smaller, manageable subtasks~\cite{correa2023humans, lovden2020human}. Inspired by this cognitive process, we design \sys{} to handle user instructions in a similar way, enabling intelligent and personalized smart home automation. Fig.~\ref{fig:architecture} shows the overall design of \sys{}. At the core is a three-stage reasoning pipeline that incrementally translates natural language instructions into executable IoT commands, such as trigger-action rules, direct control commands, or device queries\footnote{Trigger-action rules are event-driven automations (e.g., ``\textit{turn on the light when someone is in the living room}''). Direct control commands are immediate device operations (e.g., ``\textit{make the bedroom ready for sleep}''). Device queries request smart home states (e.g., ``\textit{what is the room temperature?}'').}. Each stage employs an LLM to progressively refine intermediate results, culminating in personalized outputs that effectively reflect user preferences. The three stages are as follows:
\begin{sitemize}
    \item \textbf{Decompose}: Given a user instruction, \sys{} breaks it down into several subtasks, which are concrete, device-specific operations that collectively fulfill the instruction. If the instruction is ambiguous (e.g., ``\textit{make the bedroom ready for sleep}''), \sys{} utilizes the LLM's common knowledge to propose plausible subtasks as a baseline (e.g., \{``\textit{adjust air conditioner temperature},'' ``\textit{set humidifier level},'' ``\textit{dim the sleep light}''\}). This proposal serves as a starting point rather than a final decision, acknowledging that the initial guess may not perfectly match the user's intent.

    \item \textbf{Derive}: For each subtask, \sys{} generates executable API-level device commands with abstract placeholders for unspecified parameters (e.g., ``\textit{adjust air conditioner temperature}'' $\Rightarrow$ [\textsf{\footnotesize turn on air conditioner} $\rightarrow$ \textsf{\footnotesize set mode to \textit{[mode\_value]}} $\rightarrow$ \textsf{\footnotesize set temperature to \textit{[temperature\_value]}}]). This stage focuses on producing valid command structures while keeping unspecified parameters configurable.
    
    \item \textbf{Refine}: \sys{} personalizes these commands by filling parameters based on user preferences inferred from context and past interactions. However, if preference evidence is unavailable (e.g., first-time use), \sys{} adopts conservative defaults and flags uncertain choices for user review.
\end{sitemize}
Crucially, after the Refine stage, \sys{} enters a human-in-the-loop step where users can review the proposal and, if needed, modify subtasks (e.g., adding ``\textit{play music}'') or adjust parameters to match their preferences. This design minimizes user burden by consolidating intervention into a single final review, rather than requiring step-by-step supervision. Moreover, this feedback updates both the system's learned task knowledge (i.e., \textit{task memory}) and preference profile, so that subsequent instructions better reflect user preferences. Once finalized, the resulting commands are delivered to the IoT platform (e.g., Samsung SmartThings). Trigger-action rules are installed and executed when their conditions are met, while direct control commands and device queries are executed immediately.

\newitem{Benefits of task decomposition.}
Following this architecture, our task decomposition approach provides two major benefits. First, it enables efficient reuse of knowledge across tasks, improving both efficiency and reliability. To realize this benefit, \sys{} leverages its \textit{task memory}, which records previously learned subtasks and their associated command structures. For example, once \sys{} has derived the device-level commands for the subtask ``\textit{adjust air conditioner temperature}'' while processing one instruction, it can reuse this knowledge in another instruction (e.g., ``\textit{keep the kitchen cool}'') while adapting parameters to the context (e.g., cooking) without recomputing from scratch. As the task memory grows, \sys{} becomes less dependent on repeated LLM inference, which mitigates unreliability and latency in IoT command generation.

Second, task decomposition enables fine-grained personalization at the subtask level. Since subtasks expose configurable parameters via their low-level commands, \sys{} can adjust these values based on user preferences. Moreover, user preferences learned once can be flexibly transferred to similar subtasks in future tasks. For example, if a user's preference for a ``cool'' environment has been learned through the subtask ``\textit{adjust air conditioner temperature},'' the same preference can later be applied to another subtask that also affects temperature, such as ``\textit{adjust fan speed},'' when an air conditioner is not available. This subtask-level personalization allows \sys{} to capture accumulated preferences over time and apply them to new tasks.

\subsection{Technical challenges}
While task decomposition provides clear benefits, bringing this design into practice raises several technical challenges:

\newitem{C1.} \textit{How to reliably decompose a task?} 
A key challenge in generating valid IoT commands is that LLMs often fail to consistently decompose natural language instructions into subtasks and derive their low-level device commands. This is because LLMs are prone to ambiguity and non-determinism: when asked to ``\textit{make the bedroom ready for sleep},'' they may introduce subtasks that cannot be executed in the current environment (e.g., ``\textit{adjust window blinds}'' when no smart blinds are installed), or generate invalid device commands (e.g., setting an unsupported mode on the air conditioner). Such behaviors can render the resulting IoT commands incomplete or infeasible. Therefore, ensuring reliable task decomposition requires carefully designed prompting strategies and proper correction mechanisms (discussed in Sec.~\ref{sec:pipeline}).
    
\newitem{C2.} \textit{How to efficiently support subtask reuse?}
Even when subtasks and their device commands are correctly generated, reusing them across tasks remains challenging due to contextual variation. The same subtask (e.g., ``\textit{adjust air conditioner temperature}'') may occur in different contexts, such as ``\textit{keep the kitchen cool}'' versus ``\textit{make the bedroom ready for sleep}.'' If each context-specific variant is stored separately, the task memory can quickly grow large, leading to excessive space overhead and making it difficult to reuse knowledge across similar situations. On the other hand, if subtasks are memorized without context, reuse may ignore contextual nuances and fail to properly reflect personalization. Therefore, efficient reuse requires a well-designed task memory that enables context-aware adaptation of previously stored task knowledge. This involves selectively retrieving and adapting relevant task knowledge to fit the current situation (discussed in Sec.~\ref{sec:memory}).
    
\newitem{C3.} \textit{How to adaptively personalize IoT commands?}
Conventional personalization in smart home systems relies on device interaction histories to capture user preferences, typically at the level of individual devices. However, applying these device-specific preferences poses two limitations. First, preferences captured for one device cannot be easily transferred to others with different capabilities; for example, knowing a user’s preferred air conditioner settings does not directly indicate how a fan should be configured when the fan is the only available option. Second, because preferences are context-dependent, relying solely on device usage histories makes it difficult to apply the right preferences under different contexts. To overcome these limitations, \sys{} abstracts user preferences into device-agnostic environmental properties (e.g., temperature, brightness) and maintains them separately for different contexts (e.g., \textit{sleeping} vs. \textit{exercising}). This context-aware abstraction enables flexible transfer of preferences across heterogeneous devices and contexts (discussed in Sec.~\ref{sec:personalization}).

\section{Task Decomposition Pipeline} \label{sec:pipeline}

\begin{figure}[t]               
  \centering
  \includegraphics[width=0.9\linewidth]{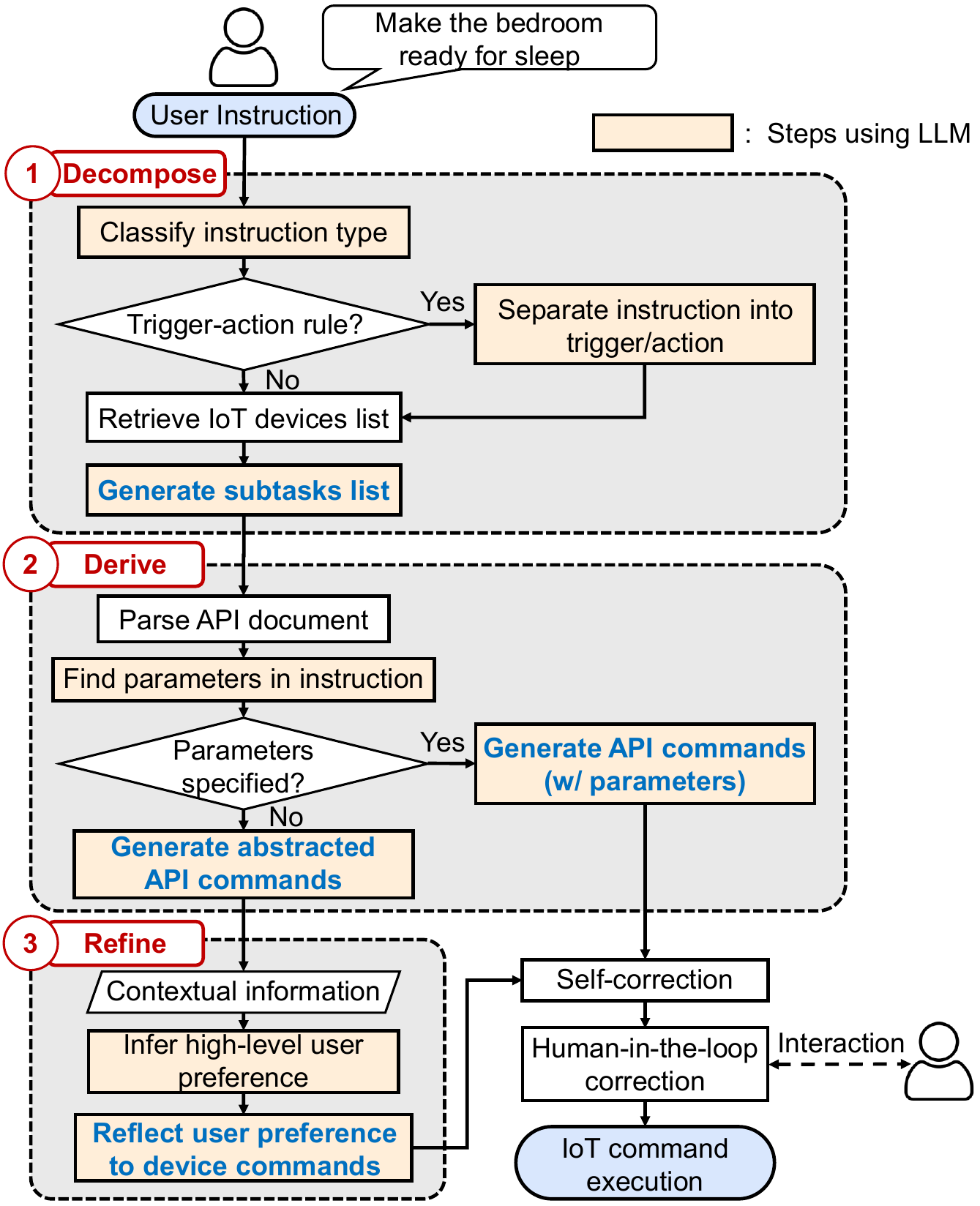}
  \caption{Workflow of the \sys{} pipeline}
  \vspace{-0.5cm}
  \label{fig:workflow}
\end{figure}

To translate natural language instructions into IoT commands, \sys{} employs the \textit{Decompose–Derive–Refine} pipeline, as illustrated in Fig.~\ref{fig:workflow}. This section outlines how the pipeline operates and how \sys{} addresses issues arising from the inherent unreliability of LLMs.

\subsection{Three Pipeline Stages}

\newitem{Decompose stage.}
The goal of the \textit{Decompose} stage is to generate a list of executable subtasks from a given user instruction. Trained on large-scale general knowledge, the LLM can infer plausible subtasks with acceptable accuracy through common-sense reasoning. To accomplish this, \sys{} first prompts the LLM to classify the instruction into one of three types: a trigger–action rule, a direct control command, or a device query. Once the type is identified, \sys{} retrieves the list of IoT devices currently connected to the environment via the IoT platform's API call. This list includes the devices along with their available functions and specifications, and is provided to the LLM as context. By grounding the reasoning process in the actual device list, \sys{} reduces the risk of introducing subtasks that cannot be executed in the current environment. Guided by the instruction type and device information, the LLM then generates the subtasks required to fulfill the instruction, along with the devices needed to execute them. In particular, for trigger–action rules, the LLM is prompted to first separate the instruction into trigger and action parts and then generate the corresponding subtasks for each part.

For example, given the instruction ``\textit{make the bedroom ready for sleep},'' which is somewhat ambiguous in specifying what concrete actions should be taken, the LLM can still leverage common-sense knowledge to infer a reasonable set of subtasks. The output of the Decompose stage would be a JSON-formatted list of subtasks such as ``\textit{adjust air conditioner temperature},'' ``\textit{set humidifier level},'' and ``\textit{dim the sleep light},'' each associated with the relevant device, as follows:

\begin{lstlisting}
{"CommandType": "Direct Control Command",
 "Action": {
  "name": "Make the bedroom ready for sleep",
  "possible subtask list":
   [{"subtask": "Adjust air conditioner temperature",
     "device": "air conditioner"},
    {"subtask": "Set humidifier level",
     "device": "humidifier"},
    {"subtask": "Dim the sleep light",
     "device": "sleep light"}]}}
\end{lstlisting}

\newitem{Derive stage.}
The goal of the \textit{Derive} stage is to translate each subtask obtained from the previous stage into a sequence of low-level device commands that strictly conform to the IoT platform's API specification. To this end, \sys{} prompts the LLM with three inputs: \textit{i)} the user instruction, \textit{ii)} the list of subtasks, and \textit{iii)} the API documentation. 
Since API documentation can be voluminous (e.g., 9.3MB for Samsung SmartThings~\cite{SmartThingsAPI}), \sys{} employs retrieval-augmented generation (RAG) pipeline that parses the API documentation and retrieves only the portions relevant to each subtask. 

When generating low-level commands, assigning appropriate parameter values is paramount. When these values are explicitly specified in the instruction, they can be directly mapped to corresponding parameters. However, user instructions are often ambiguous and lack details. Continuously requesting users for missing information would incur excessive interaction overhead. Instead, \sys{} leaves values as abstract placeholders, which are later filled during the Refine stage to reflect user preferences and device context.  

Continuing the earlier example, the subtask ``\textit{adjust air conditioner temperature}'' may be converted into the following command sequence: [\textsf{\footnotesize turn on air conditioner} $\rightarrow$ \textsf{\footnotesize set mode to \textit{[mode\_value]}} $\rightarrow$ \textsf{\footnotesize set temperature to \textit{[temperature\_value]}}]. Since the parameter values are not specified in the original instruction, they are represented as abstract placeholders.

\begin{lstlisting}
{"subtask": "Adjust air conditioner temperature",
 "commands":
 [{"desc": "Turn on air conditioner",
   "device": {"name": "air conditioner",
     "capability": {"name": "switch",
     "command": "on", "value":{}}}},
  {"desc": "Set mode to [mode_value]",
   "device": {"name": "air conditioner",
     "capability": {"name": "airConditionerMode",
     "command": "setAirConditionerMode",
     "value":{"string": "[mode_value]"}}}},
  {"desc": "Set temperature to [temperature_value]",
   "device": {"name": "air conditioner",
     "capability":
     {"name": "thermostatCoolingSetpoint",
      "command": "setCoolingSetpoint",
      "value":{"decimal": "[temperature_value]"}}}}]}
\end{lstlisting}

\newitem{Refine stage.}
The goal of the \textit{Refine} stage is to personalize the low-level commands to align with user preferences. To do this, \sys{} prompts the LLM with \textit{i)} the subtasks and their low-level commands, and \textit{ii)} a structured summary of user preferences. These preferences are extracted from past device interactions and represented at the level of environmental properties (e.g., temperature, brightness, noise). Based on these preferences, the parameter values of the abstracted commands are refined. For instance, parameter slots in commands like [\textsf{\footnotesize set temperature to \textit{[temperature\_value]}}] are filled with personalized values (e.g., 20°C), or default values if context-specific preferences are unavailable. A more detailed description of this personalization process is provided in Sec.~\ref{sec:personalization}.

\subsection{Two-Step Correction Process}
Even with a structured pipeline, \sys{} can still produce imperfect outputs. First, generated commands may exhibit inconsistent or erroneous behavior (e.g., API-level errors) due to the stochastic nature of LLMs. Second, personalization can be under-specified: for a given context, the system may lack sufficient preference evidence, leading to conservative defaults or mismatched parameter choices. To address both issues, \sys{} adopts a two-step correction process that combines automated validation with optional user intervention.

\newitem{Automatic self-correction.}
\sys{} first attempts automated self-correction using a virtual smart home environment composed of simulated devices mirroring the real deployment. Before applying commands to the physical environment, \sys{} executes them in the virtual environment and inspects the resulting execution logs for API violations and runtime failures. When an error is detected, \sys{} feeds the error message back to the LLM and prompts it to revise only the problematic parts of the command sequence. This log-driven revision repeats until a retry limit is reached (e.g., 3–5 rounds), providing a lightweight safeguard without user involvement.

\newitem{Human-in-the-loop correction.}
If errors persist after self-correction, or if the user wants to override ambiguous personalization choices, \sys{} escalates to an interactive correction step. The LLM presents a brief explanation of the generated commands, allowing users to intervene via natural-language feedback. Users can specify corrections to fix remaining validity issues (e.g., removing erroneous subtasks) or further personalize the automation by adjusting parameters or adding new subtasks. In the latter case, \sys{} re-invokes the pipeline from the Derive stage to integrate the new requirements. This feedback corrects the current task while updating the system's task memory and preference profile for future instructions.

\section{Hierarchical Task Memory} \label{sec:memory}

\begin{figure*}[t]               
  \centering
  \includegraphics[width=0.85\linewidth]{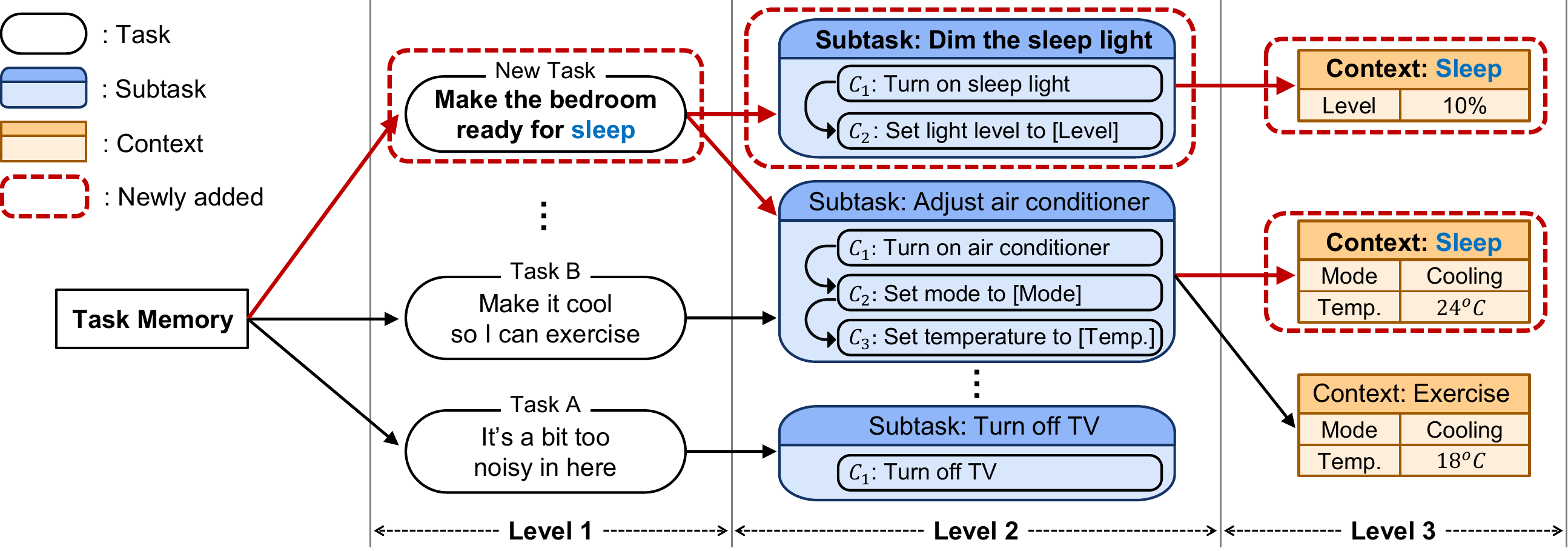}
  \caption{Task memory structure of \sys{}}
  \vspace{-0.5cm}
  \label{fig:memory}
\end{figure*}

The task knowledge produced by the three-stage pipeline is systematically archived in \sys{}'s task memory. This memory serves as a knowledge base that allows \sys{} to make subsequent IoT command generation more reliable and efficient by reusing previously learned subtasks and device commands. In this section, we describe how \sys{} organizes and stores the pipeline outcomes, and how this knowledge is recalled for future IoT command generation.

\subsection{Task Memory Structure}
As illustrated in Fig.~\ref{fig:memory}, \sys{} organizes its task memory as a hierarchical directed acyclic graph (DAG) with three depth levels: \textit{task} (level 1), \textit{subtask} (level 2), and \textit{context} (level 3). Each level captures knowledge at a different granularity, enabling both reuse and personalization.

\newitem{Task-level knowledge.} Each task node corresponds to a user instruction issued in natural language. When a new instruction arrives, \sys{} first checks---before performing the Decompose stage---whether a similar task already exists in the memory by comparing the incoming instruction against stored tasks through text-embedding-based cosine similarity. If a similar task exists, the stored knowledge can be reused (as described later in the recall process); otherwise, a new task node is created and the Decompose stage is performed.

\newitem{Subtask-level knowledge.}
Subtask nodes are created during the Decompose stage, representing the subtask operations extracted from a task. Subtask nodes are linked to their originating task node. Importantly, subtasks are not restricted to a single task but can be shared across multiple tasks. For instance, when a new subtask is generated from the Decompose stage, \sys{} checks for similarity against existing subtask nodes in the memory; if a match is found, it simply adds a new edge between the current task node and the matched subtask node. In this way, previously stored subtasks can be directly reused across different tasks. If a match is not found, a new subtask node is created. 

Each subtask node stores the sequence of low-level device commands generated during the Derive stage. When parameter values are explicitly specified in the instruction and appear as concrete values in the generated commands, \sys{} replaces them with abstracted placeholders (e.g., \textsf{\footnotesize [temperature\_value]}) before storing the commands in task memory. This generalization preserves the structure of the command sequence while allowing the parameter slots to be filled later.

\newitem{Context-level knowledge}:
Context nodes are added during the Refine stage to capture how command parameters are personalized under different instruction contexts (i.e., instruction's goal or intention). For instance, when processing the instruction ``\textit{make the bedroom ready for sleep},'' \sys{} uses the LLM to extract the keyword ``\textit{sleeping}'', which represents the instruction context, and creates a corresponding node. This node stores the personalized parameter values (e.g., \textsf{\footnotesize [mode\_value]} = ``\textit{cooling},'' \textsf{\footnotesize [temperature\_value]} = ``\textit{20°C}''). If a node with the same keyword already exists, \sys{} can reuse the stored values directly.

Moreover, updates from the human-in-the-loop correction are also reflected in task memory. When users add/remove subtasks or fine-tune parameters, the corresponding subtask or context nodes are updated accordingly. As a result, corrections are not only applied to the current execution but also accumulated for future reuse, so that task memory ultimately preserves user-validated knowledge.

\subsection{Task Knowledge Recall}
When recalling task knowledge, \sys{} must account for the fact that the environment or user context may differ from when the knowledge was originally memorized. For example, some devices may no longer be available due to failures, or the same subtask may appear under a different instruction context. Therefore, it is important to determine up to which depth level knowledge should be reused, depending on the situation. When a new instruction arrives, \sys{} proceeds as follows:

\newitem{Task-/subtask-level reuse.}
If a matching task exists in the memory, or if some subtasks obtained match stored subtasks, \sys{} reuses the corresponding subtask knowledge. In such cases, the Decompose stage for the task or the Derive stage for those subtasks can be skipped. However, if certain devices associated with subtasks are unavailable (e.g., due to malfunction), those subtasks are excluded. \sys{} then notifies the user and suggests alternative subtasks through the LLM.

\newitem{Context-level reuse}:
For reusable subtasks, \sys{} checks whether relevant context knowledge can also be recalled by comparing context keywords. If a matching context node exists, the system combines the abstracted low-level commands from the subtask node with the parameter values from the context node, thereby avoiding redundant personalization. Otherwise, \sys{} reuses only the subtask-level knowledge at depth level 2, retaining abstracted parameter slots. These slots are then filled through personalization during the Refine stage.

By leveraging this memory recall mechanism, \sys{} minimizes reliance on repeated LLM inference, achieving more consistent and efficient IoT command generation.

\section{Adaptive Personalization} \label{sec:personalization}

\begin{figure*}[t]               
  \centering
  \includegraphics[width=0.85\linewidth]{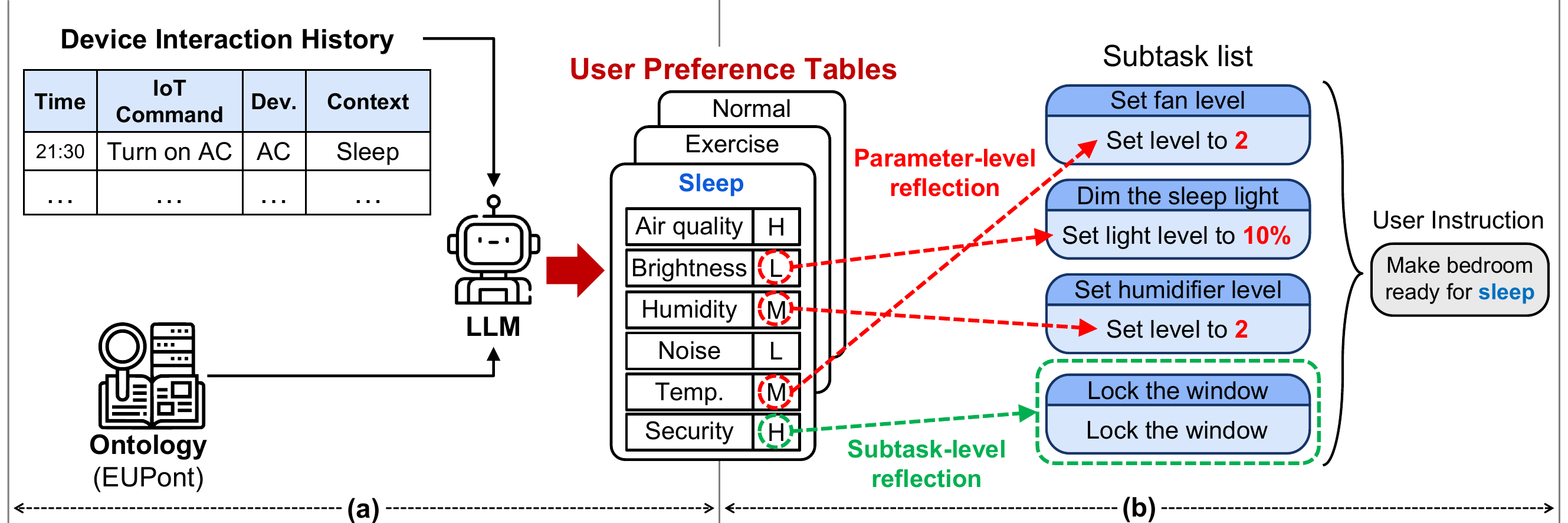}
  \caption{Adaptive personalization of \sys{}}
  \vspace{-0.5cm}
  \label{fig:personalization}
\end{figure*}

Personalization is indispensable for smart home automation, as even the same subtask may require different parameter values depending on a user's preferences and context. However, conventional approaches typically rely on device-specific interaction histories, which makes it difficult to extend personalization to other devices without prior usage data. To overcome this limitation, \sys{} abstracts user preferences into device-agnostic environmental properties (e.g., temperature, brightness). Such high-level representation enables flexible transfer of preferences across heterogeneous devices. This section describes how \sys{} extracts user preferences in terms of environmental properties and applies them to personalize IoT commands during the Refine stage.

\subsection{User Preference Extraction} 
As illustrated in Fig.~\ref{fig:personalization}(a), \sys{} prompts the LLM with \textit{i)} device interaction histories and \textit{ii)} relationships between IoT commands and environmental properties to extract user preferences in a high-level representation. To capture these relationships, we leverage the EUPont ontology~\cite{corno2017high}, which defines six environmental properties that influence the smart home environment: \textit{air quality, brightness, humidity, noise, temperature,} and \textit{security}. In particular, this ontology specifies which properties each IoT command affects and whether those properties increase or decrease. For example, adjusting the air conditioner temperature has the main effect of lowering the temperature. By supplying this ontology knowledge as guidance, \sys{} enables the LLM to infer the effects of IoT commands in the device history. However, preference extraction raises several issues:

\newitem{Context-dependent preferences.} 
User preferences often vary by context. For example, a user's preferred temperature while sleeping may differ significantly from their preference while exercising. To capture such variations, our device interaction history includes not only raw device commands but also the instruction context (i.e., instruction's goal or intention). This allows the LLM to extract separate preferences under diverse contexts such as \textit{sleeping}, \textit{exercising}, or \textit{normal}. Here, normal denotes overall preferences without context distinction.
    
\newitem{Defining property levels.}
Preferences inferred from past IoT commands must be expressed as levels for each environmental property. A straightforward approach is to partition device parameter ranges into discrete levels (e.g., \textit{low}, \textit{medium}, \textit{high}) using simple rule-based heuristics. However, defining appropriate rules for partitioning is challenging because heterogeneous devices expose diverse parameter ranges (e.g., temperature in °C, humidity in \%, brightness in lux), and discrete level boundaries can differ across properties. Moreover, for properties like \textit{security}, levels are not determined by parameter values but by broader usage patterns such as the frequency of locking windows or the duration of home camera usage. Handcrafted rules cannot capture such diversity and context dependence. To address this, \sys{} leverages the LLM with few-shot prompting, providing examples of property–level mappings (e.g., temperature, humidity, security). This enables the LLM to flexibly partition parameter ranges or interpret usage patterns into appropriate property levels.

Through this process, \sys{} builds tables mapping environmental properties to preferred levels (e.g., \textit{temperature $\rightarrow$ low}). These tables are maintained separately for different contexts (e.g., \textit{sleeping}, \textit{exercising}), enabling \sys{} to capture fine-grained variations in user preferences across situations. To keep these mappings up to date, the extraction process runs periodically (e.g., daily) or on demand during the Refine stage when preferences become stale.

\subsection{User Preference Reflection}
As shown in Fig.~\ref{fig:personalization}(b), during the Refine stage, \sys{} prompts the LLM with three inputs: \textit{i)} the instruction, \textit{ii)} the subtasks, and \textit{iii)} a preference table selected by context. To select the table, \sys{} checks whether the instruction's context keyword matches any keyword defined in the preference tables. If a match is found, the corresponding table is used; otherwise, the system defaults to the \textit{normal} context preferences. \sys{} then reflects the preferences in two ways:

\newitem{Parameter-level reflection.}
At this level, the LLM adjusts the parameter values of the given subtasks by referring to the selected preference table. For example, in a sleep-environment scenario where only a fan is available instead of an air conditioner, \sys{} can still adapt the fan's speed to match the preferred temperature level specified in the table, even if no prior usage history exists for the fan. As mentioned in Sec.~\ref{sec:memory}, the personalized parameter values are then stored in the context nodes of the task memory.

\newitem{Subtask-level reflection.}
Beyond adjusting parameters, \sys{} also introduces additional subtasks guided by user preferences. For instance, if a user exhibits a strong preference for the \textit{security} property, the LLM may add a new subtask such as ``lock the windows,'' which was not considered during the Decompose stage. This new subtask then proceeds through the Derive stage to generate its commands and, if needed, undergoes parameter-level reflection. In this way, preference reflection complements the Decompose stage by injecting user-specific adjustments that would otherwise be overlooked.

It is worth noting that when no context-specific preference table matches the instruction, \sys{} falls back to the normal preferences as a baseline. Over time, however, users can fine-tune parameter values in different contexts through \sys{}'s human-in-the-loop correction. These adjustments generate additional interaction logs, which are fed into the preference extraction process. The process then updates existing tables or creates new context-specific ones, allowing them to gradually diverge from the normal context.

\section{Implementation Notes}

We implemented a prototype of \sys{} consisting of an Android mobile app and a backend server connected over Wi-Fi. The mobile app serves as the user-facing interface: it receives natural-language instructions and forwards them to the backend. The backend manages the task memory and executes the three-stage reasoning pipeline (Decompose–Derive–Refine) to process each instruction. 
For reasoning, we designed \sys{} with a modular interface to the backbone LLM. By default, our prototype uses GPT-4o~\cite{GPT-4o}, but the backbone model can be replaced with other LLMs without changing the rest of the system. We also set the decoding temperature to 0.0 to ensure consistent and deterministic outputs.
For task knowledge retrieval, we implemented a text-embedding–based similarity search using the \texttt{text-embedding-3-small} model~\cite{OpenAIEmbedding}, applying cosine similarity between stored nodes and the current input to identify reusable subtasks or contexts. Finally, the backend interacts with the smart home environment through the Samsung SmartThings API~\cite{SmartThingsAPI}, while the IoT communication layer is modularized to allow easy extension of \sys{} to other ecosystems such as Google Home or Apple HomeKit.
Our prototype was deployed on a Google Pixel 6 smartphone as the client device, with a backend server equipped with an Intel i7-13700F CPU and 32~GB RAM.

\section{Performance Evaluation} \label{sec:performance_test}

We conducted a quantitative evaluation of \sys{} to assess its ability to translate natural-language instructions into IoT commands. We compared \sys{} against existing LLM-based smart home agents in terms of \textit{accuracy}, \textit{latency}, and \textit{cost}. To enable controlled testing over a diverse command space, we built a virtual smart home using the SmartThings Edge Driver~\cite{SmartThingsEdgeDriver}, simulating a diverse set of 26 devices across categories such as lighting, heating, and air conditioning.

\subsection{Comparative Study}

\subsubsection{Experimental Setup}
For the comparative study, we configured the experimental environment as follows:

\newitem{Baselines.}
We compared \sys{} with two LLM-based smart home agents: \sasha{}~\cite{Sasha24} and \sage{}~\cite{SAGE25}. \sasha{} is one of the earliest LLM-based smart home agents that employs a reasoning pipeline but executes each instruction as a single step without task decomposition. \sage{}, the state-of-the-art framework at the time of our experiments, adopts a dynamic approach in which the LLM plans the execution order of reasoning modules at runtime. While this allows flexible decomposition, \sage{} does not memorize knowledge from prior executions; instead, it replans from scratch for every instruction. We utilized the open-source version~\cite{SAGE_code} of \sage{}. Since \sasha{}'s source code is unavailable, we reimplemented it following the pipeline architecture detailed in its original paper.

\newitem{Dataset.}
We used datasets from \sasha{} and \sage{}. Each task in these datasets consists of a natural-language instruction paired with ground-truth device commands labeled by human annotators. We excluded tasks incompatible with the SmartThings platform. Furthermore, because the original datasets contain relatively few trigger-action cases, we created 15 additional tasks. Consequently, our final dataset consists of 97 tasks: 35 from \sasha{}, 47 from \sage{}, and 15 newly added by us. These tasks span three categories: 55 direct control commands, 32 trigger-action rules, and 10 device queries.

\newitem{Metrics.}
To rigorously evaluate accuracy, we defined four complementary metrics. The \textit{successful task rate (STR)} is the proportion of tasks where all required ground-truth commands are correctly generated and executed (ignoring parameter personalization). The \textit{excessive control rate (ECR)} is the proportion of tasks containing extra commands not in the ground truth, while the \textit{insufficient control rate (ICR)} is the proportion of tasks missing at least one required command. The \textit{syntax error rate (SER)} is the proportion of tasks containing at least one command that violates the SmartThings API specifications (e.g., syntax errors). Note that ECR, ICR, and SER are not mutually exclusive, as a task may simultaneously include unnecessary commands, omit required ones, and contain syntax errors. In terms of efficiency, we also measured \textit{latency} (end-to-end time from instruction issuance to completion of command execution in the virtual smart home) and \textit{cost} (USD per instruction computed from aggregated input/output tokens across all model calls using the API provider’s pricing).

\newitem{Procedure.}
We evaluated \sys{} in two phases: \syscold{} and \syswarm{}. In \syscold{}, the agent processed all instructions for the first time while building task memory and reused knowledge as soon as it became available. The \syswarm{} phase then utilized the fully populated task memory to reprocess the same tasks, but with a rephrased version of the dataset to prevent reliance on trivial string matching. Specifically, we rephrased natural language instructions using GPT-4o while preserving semantic intent. This compelled the system to leverage knowledge based on semantic understanding. In contrast, \sasha{} and \sage{} processed each instruction only once without phase distinction. To ensure fair comparison, we disabled all user-feedback steps and applied the same backbone model across all approaches within each run. Furthermore, to verify generalizability, we evaluated using four distinct LLMs (GPT-4o~\cite{GPT-4o}, GPT-5.2~\cite{GPT-5.2}, Gemini 3 flash~\cite{gemini3-flash}, and Claude Opus 4.5~\cite{claude4.5-opus}), averaging results over three runs.

\begin{figure*}[t]
    \centering
    \includegraphics[width=\linewidth]{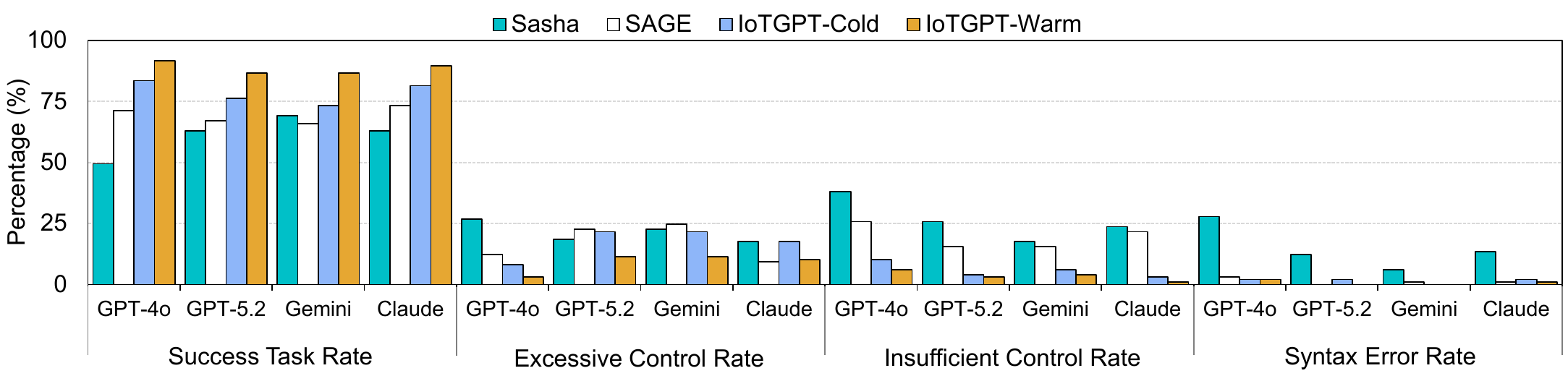}
    \caption{Average accuracy of \sasha{}, \sage{} and \sys{}}
    \vspace{-0.5cm}
    \label{fig:comparison_test}
\end{figure*}



\subsubsection{Accuracy}
Fig.~\ref{fig:comparison_test} presents four accuracy metrics for the three approaches across four LLM backbones. Except for Gemini 3 flash, \sasha{} consistently shows the lowest STR across models (e.g., 49.48\% on GPT-4o). Because it processes instructions monolithically without fine-grained decomposition, \sasha{} often misses essential device commands and produces invalid syntax under ambiguous or complex tasks, resulting in the highest ICR and SER among the agents. \sage{} generally improves STR over \sasha{} by leveraging more advanced reasoning modules and dynamically planning their execution order. However, its reliance on the LLM often produces long planning chains, which increases error accumulation due to the model’s stochasticity. Consequently, \sage{} still exhibits a relatively high ICR, with some required commands missed and tasks left incomplete.

In contrast, \syscold{} achieves higher STR than both \sasha{} and \sage{}, reaching up to 83.51\% across the four backbones. It also yields lower ICR and SER, indicating fewer missing commands and syntax errors. However, \syscold{} shows a non-negligible ECR across models, ranging from 8.25\% to 21.65\%. Our inspection revealed that many of these ``excessive'' commands were not harmful errors but potentially useful additions. For instance, when processing the task ``\textit{If there's no one at home, save energy},'' \syscold{} generates not only the ground-truth commands to turn off appliances but also an additional door lock command. Although this command is not included in the ground truth, some users might consider it essential for fulfilling the task. Such outputs are penalized under our strict ECR definition, but they highlight \syscold{}'s potential to provide richer support tailored to user needs.

Crucially, the accuracy of \sys{} improves further in the \syswarm{} phase, where task memory is leveraged to reuse previously learned knowledge. Specifically, the STR increases by up to 18.31\% compared to \syscold{} across the evaluated models, directly attributable to the reduced reliance on repeated LLM inference. These results suggest that as task memory grows, \sys{} can continue to improve by accumulating reusable knowledge.

Interestingly, we observe distinct backbone sensitivities across systems. Because \sasha{} relies on a relatively simple prompt design, its performance is highly sensitive to the backbone model's capabilities. For instance, under Gemini 3 flash, \sasha{} outperforms \sage{} in STR, largely because enhanced reasoning substantially reduces syntax errors. However, despite this boost, \sasha{} still records the highest ICR, indicating that monolithic inference remains prone to missing required commands. In contrast, \sys{} exhibits a counter-intuitive STR decline with more capable models like GPT-5.2 and Gemini 3 flash. Paradoxically, this stems from their advanced reasoning, which tends to generate a larger number of additional, context-aware subtasks; while potentially useful, these additions are absent from the fixed ground truth, thereby inflating ECR and lowering STR.

\begin{figure}[t]
\vspace{-0.5cm}
\centering
\subfloat[ ]{%
  \includegraphics[width=0.49\columnwidth]{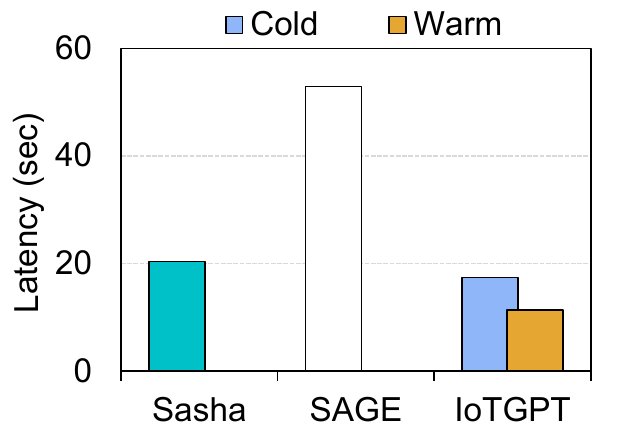}%
}\hfil
\subfloat[  ]{%
  \includegraphics[width=0.49\columnwidth]{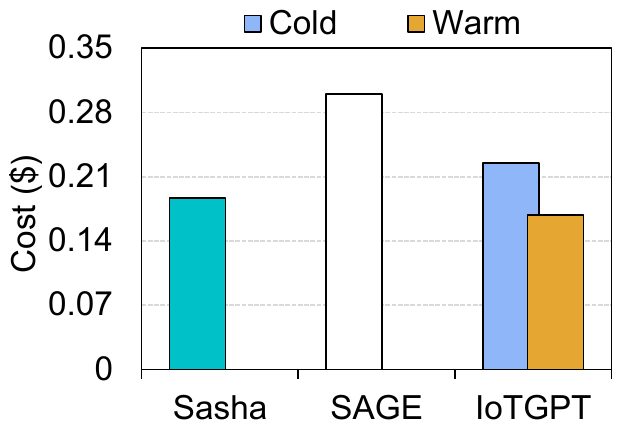}%
}
\caption{(a) Average latency of \sasha{}, \sage{}, and \sys{}. (b) Average cost of \sasha{}, \sage{}, and \sys{}.}
\vspace{-0.5cm}
\label{fig:comparison_test_latency_cost}
\end{figure}

\subsubsection{Latency \& Cost}
Fig.~6 illustrates the average latency and cost per instruction for each approach using the GPT-4o backbone. \sage{} exhibits the worst performance across both metrics due to its heavy reliance on the LLM for dynamically planning reasoning modules. While this flexibility can occasionally improve reasoning quality, it introduces substantial overhead, particularly for complex tasks where planning chains become excessively long. Indeed, we observed some instructions taking up to 151 seconds, and the per-instruction cost reaching as high as \$4.2 (e.g., ``\textit{If there is no one at home, save energy}''). In contrast, \sasha{} shows relatively lower latency and cost than \sage{}, largely due to its simple monolithic pipeline that processes instructions directly without decomposition. However, this simplicity comes at the expense of accuracy, as shown in our accuracy results.

Meanwhile, \syscold{} presents an interesting trade-off. Although it incurs higher cost than \sasha{} due to the cumulative token usage of its multi-stage pipeline, it achieves lower latency. We attribute this counter-intuitive result to the use of more structured and detailed prompts at each stage, which can reduce the LLM's inference time despite multiple stages.

Furthermore, \syswarm{} demonstrates significant efficiency gains by leveraging task memory, achieving the lowest latency and cost among all approaches. Compared to \syscold{}, latency and cost decrease by 34.44\% and 25.33\%, respectively. This reduction stems from reusing learned subtasks without repeated LLM inference. Overall, these results show that \sys{} becomes progressively more efficient as its task memory accumulates reusable knowledge.

\subsection{Ablation Study}
We conducted an ablation study to analyze the contributions of \sys{}'s key features. Specifically, we removed task decomposition and task memory to assess their impact on accuracy, latency, and cost. Note that personalization was excluded from this ablation study, as it relies on subjective user experience and is evaluated separately via the user study in Sec.~\ref{sec:personalization_eval}.

\subsubsection{Experimental Setup}
For the ablation study, we configured the experimental environment as follows:

\newitem{Baselines.}
We compared \sys{} against two custom baselines: \nodecomp{} and \nomem{}. \nodecomp{} removes task decomposition and directly infers the required device commands from a user instruction, following the derive-only design used in prior agents~\cite{Sasha24, RuleBot++24, AwareAuto24}. \nomem{} retains task decomposition but disables the task memory, executing the pipeline without reusing knowledge from past subtasks. For a fair comparison, both baselines share the same prompts as \sys{}.

\newitem{Dataset \& Metrics.}
We used the same dataset as in the comparative study, and measured STR as the primary accuracy metric, alongside latency and cost.

\newitem{Procedure.}
We evaluate the ablation study in cold-start and warm-start stages, employing GPT-4o as the backbone model for all agents. In cold-start, each agent processed all instructions for the first time. During this stage, both \sys{} and \nodecomp{} built task memory and reused available knowledge on the fly. The key distinction lies in granularity: \sys{} stored knowledge at the fine-grained subtask level, whereas \nodecomp{} stored it only at the task level due to the absence of decomposition. In warm-start, each agent reprocessed the same tasks using a rephrased version of the dataset, as in the comparative study. \sys{} and \nodecomp{} leveraged their constructed memory, while \nomem{} executed the reasoning pipeline from scratch for every instruction.

\begin{figure*}[t]
\centering
\subfloat[]{%
  \includegraphics[width=0.28\textwidth]{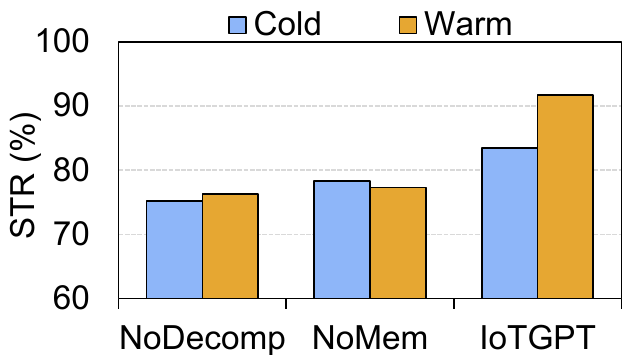}
}\hfil
\subfloat[]{%
  \includegraphics[width=0.28\textwidth]{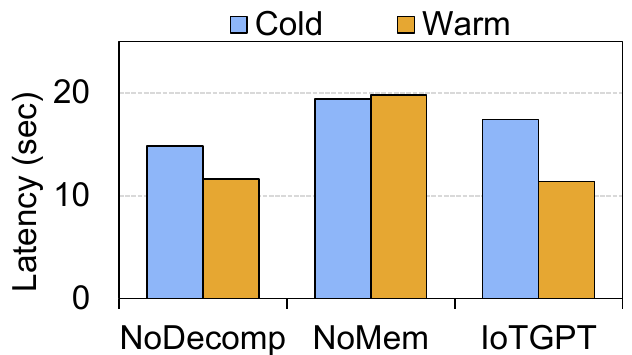}
}\hfil
\subfloat[]{%
  \includegraphics[width=0.28\textwidth]{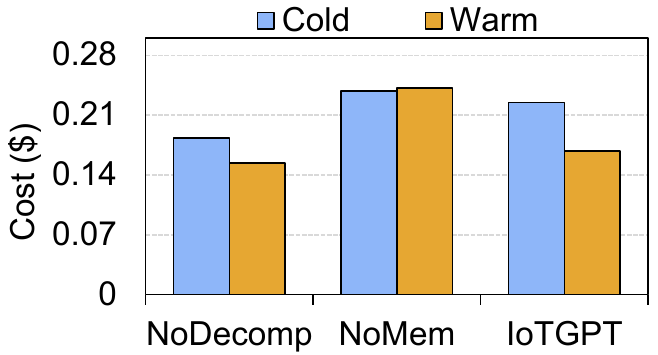}
}
\caption{Ablation results of \sys{} under cold-start (Cold) and warm-start (Warm) settings}
\vspace{-0.5cm}
\label{fig:ablation_study}
\end{figure*}

\subsubsection{Accuracy}
Fig.~\ref{fig:ablation_study}(a) shows STR for each approach. We analyze the impact of task decomposition by comparing \sys{} with \nodecomp{}. During cold-start, \sys{} outperforms \nodecomp{}, demonstrating that decomposing tasks into subtasks improves reasoning accuracy. This performance gap widens in the warm-start stage. This is largely because \nodecomp{} relies on coarse-grained task-level memory and often fails to retrieve reusable knowledge when rephrased instructions exhibit high linguistic variation, yielding only a marginal STR improvement of 1.37\%. In contrast, \sys{} leverages fine-grained subtask-level reuse; even if task-level matching fails due to rephrasing, it can reuse knowledge for decomposed subtasks. This granular reuse enables \sys{} to benefit from learned knowledge across tasks when they share common derived subtasks, leading to a substantial 9.87\% improvement. This implies that task decomposition not only improves reasoning performance but also synergizes with task memory by enabling robust reuse under linguistic variations.

Next, we examine the role of task memory by comparing \sys{} with \nomem{}. During cold-start, \sys{} builds task memory from successful executions and reuses available knowledge instead of repeatedly relying on the LLM, thereby reducing opportunities for command-generation errors. Conversely, \nomem{}, which relies entirely on the LLM without memory, exhibited a lower STR due to frequent errors. Notably, in warm-start, \nomem{} shows a 1.31\% drop in STR on the rephrased dataset, despite reprocessing tasks that are semantically similar to those it has seen before. This observation highlights that task memory plays an important role in reducing errors in command generation.

\subsubsection{Latency \& Cost}
Fig.~\ref{fig:ablation_study}(b) and \ref{fig:ablation_study}(c) show the latency and cost of each approach, respectively. When comparing \sys{} with \nodecomp{}, \sys{} incurs higher overhead during the cold-start stage. This increase stems primarily from the inclusion of detailed device specifications in the prompt required for task decomposition, representing a necessary trade-off for enhanced accuracy. However, in the warm-start stage, \sys{} achieves efficiency comparable to \nodecomp{} by leveraging task memory to bypass redundant LLM inference. In contrast, \nomem{} consistently exhibits the highest latency and cost across both stages, as it is compelled to repeat the decomposition process for every instruction. These results suggest that while task decomposition introduces inherent overhead, task memory significantly mitigates this burden, ensuring system efficiency.

\section{User Studies} \label{sec:user_studies}

To complement the quantitative evaluation in Sec.~\ref{sec:performance_test}, we conducted two user studies: \textit{i) Study 1: Personalization under changing smart home environments}, which examined whether \sys{} can adapt user preferences when device configurations change; and \textit{ii) Study 2: Overall usability across task difficulty levels}, which evaluated the usability of \sys{} compared to baseline systems for automation tasks of varying complexity.

We recruited 15 participants (11 male, 4 female; mean age = 28.1, range = 24–32) through online community advertisements. Four participants had prior experience with IoT platforms such as Google Home or Samsung SmartThings, and 14 had experience with conversational agents including Google Assistant, Apple Siri, or Samsung Bixby. Each session lasted 60–80 minutes, and participants received 10~USD compensation.
The user studies were in accordance with our institution's IRB policies and the consent forms.

\begin{table}[t]
\caption{Device Configurations of the \baseenv{} and \newenv{} Smart Home Environments Used for the Personalization Study.\label{tab:smart_home_settings}}
\centering
\begin{tabularx}{\columnwidth}{|c|Y|Y|}
\hline
 & \textbf{BASE env.} & \textbf{NEW env.} \\
\hline
Temperature & Air conditioner, Thermostat, Temp/Humidity sensor & Temp/Humidity sensor, Fan \\
\hline
Humidity & Air conditioner & Fan \\
\hline
Brightness & Light, Blind & Light, Blind \\
\hline
Noise & TV, Vacuum cleaner & TV, Vacuum cleaner \\
\hline
Air quality & Air conditioner, Air-quality sensor & Smart window, Air-quality sensor \\
\hline
Security & Door lock & Home camera \\
\hline
\end{tabularx}
\vspace{-0.5cm}
\end{table}

\subsection{Study 1: Adaptive Personalization Evaluation}
\label{sec:personalization_eval}

We conducted a user study to assess \sys{}'s personalization. Unlike quantitative metrics, whether generated parameters reflect user preferences is inherently subjective. Thus, we relied on human evaluation to examine whether \sys{} effectively personalizes parameters as device availability changes.

\subsubsection{Experimental Setup} 
We set up our experimental environments as follows:

{\begin{table*}[t]
    \centering
    \caption{Task Set Used in the Personalization Study. Tasks Are Grouped into Three Difficulty Levels Based on the Number of Environmental Properties Involved. B = Brightness, T = Temperature, H = Humidity, N = Noise, A = Air Quality, S = Security.}
    \label{tab:preference_study_dataset}
    \begin{tabular}{>{\centering\arraybackslash}m{3.0cm}|>{\arraybackslash}m{6.8cm}|>{\centering\arraybackslash}m{0.8cm}|>{\centering\arraybackslash}m{0.8cm}|>{\centering\arraybackslash}m{0.8cm}|>{\centering\arraybackslash}m{0.8cm}|>{\centering\arraybackslash}m{0.8cm}|>{\centering\arraybackslash}m{0.8cm}}
        \hline
        \textbf{Difficulty level} & \textbf{Tasks} & \multicolumn{6}{c}{\textbf{Env. properties}} \\
        \cline{3-8} \textbf{(\# of properties)} & \textbf{(user instructions)} & B & T & H & N & A & S \\
        \hline
        \multirow{3}{*}{Easy (1–2)} 
        & Help me cool off& &  \checkmark&  & &  &  \\
        & Change the mood to something that lifts my spirits & \checkmark &  &  & \checkmark &  &  \\
        & It’s a bit too noisy in here &  &  &  & \checkmark &  &  \\
        \hline
        \multirow{3}{*}{Medium (3–4)} 
        & Set up a good study environment & \checkmark & \checkmark & \checkmark & \checkmark &  &  \\
        & Set the room for a relaxing atmosphere & \checkmark & \checkmark &  & \checkmark &  &  \\
        & I'm tired – make it cozy so I can rest & \checkmark & \checkmark &  & \checkmark &  &  \\
        \hline
        \multirow{2}{*}{Hard (5–6)} 
        & Make the house a comfortable atmosphere & \checkmark & \checkmark & \checkmark & \checkmark & \checkmark &  \\
        & Make me the best sleeping environment & \checkmark & \checkmark & \checkmark & \checkmark & \checkmark & \checkmark \\
        \hline
    \end{tabular}
\end{table*}}

{\begin{table*}
    \centering
    \setlength{\tabcolsep}{3pt}
    \caption{Survey Results from Study 1. Each Rating Was Collected on a 7-Point Likert Scale after Task Completion. Values Represent Means, with Standard Deviations Shown in Parentheses.}
    \label{tab:preference_study_result}
    \begin{tabular}{p{7cm}
    |>{\centering\arraybackslash}m{1.6cm}
    >{\centering\arraybackslash}m{1.6cm}
    >{\centering\arraybackslash}m{1.6cm}
    |>{\centering\arraybackslash}m{1.6cm}
    >{\centering\arraybackslash}m{1.6cm}
    >{\centering\arraybackslash}m{1.6cm}}
        \hline
        \multirow{2}{*}{\rule{0pt}{1.0\normalbaselineskip}} & \multicolumn{3}{c|}{\textbf{BASE env.}} & \multicolumn{3}{c}{\textbf{NEW env.}} \\
        \cline{2-7}
         & \rule{0pt}{1.0\normalbaselineskip}Sasha & SAGE & IoTGPT & Sasha & SAGE & IoTGPT \\
        \hline

        \makecell[l]{\textbf{Q1.} Were all necessary device commands included \\ (regardless of preferences)?}
        & 2.47 $\pm$ 0.21 & 3.68 $\pm$ 0.13 & 5.95 $\pm$ 0.05 & 1.93 $\pm$ 0.19 & 3.90 $\pm$ 0.13 & 5.78 $\pm$ 0.05 \\
        \hline
\makecell[l]{\textbf{Q2.} Were the selected devices appropriate for the \\ preferences?}
        & 2.44 $\pm$ 0.23 & 3.52 $\pm$ 0.14 & 6.28 $\pm$ 0.04 & 1.93 $\pm$ 0.19 & 3.72 $\pm$ 0.14 & 5.63 $\pm$ 0.09 \\
        \hline
\makecell[l]{\textbf{Q3.} Were the device settings adjusted according to \\ the preferences?}
        & 2.20 $\pm$ 0.21 & 3.04 $\pm$ 0.11 & 6.02 $\pm$ 0.05 & 1.93 $\pm$ 0.19 & 3.18 $\pm$ 0.14 & 5.57 $\pm$ 0.10 \\
        \hline
    \end{tabular}
    \vspace{-0.3cm}
\end{table*}}

\newitem{Smart home settings.}
As illustrated in Table~\ref{tab:smart_home_settings}, we prepared two environments, \baseenv{} and \newenv{}, each with nine devices. Under the EUPont ontology, each device is associated with specific environmental properties. While several devices are shared, the two environments differ in a subset of devices. For instance, \baseenv{} includes an air conditioner, thermostat, and smart door lock, whereas \newenv{} incorporates a fan, smart window, and home camera. This setup simulates heterogeneous smart home setups, challenging the system to adapt user preferences to available device capabilities.

\newitem{Approaches to be compared.}
We compared the same approaches as in the previous experiments, namely \sasha{} and \sage{}. For \sys{}, we did not evaluate it separately in cold-start and warm-start, since the presence or absence of task memory had no meaningful effect on personalization quality. We used GPT-4o as the backbone model for all approaches. 

\newitem{Dataset.}
As shown in Table~\ref{tab:preference_study_dataset}, we selected 8 tasks from the dataset introduced in Sec.~\ref{sec:performance_test}. Tasks were chosen based on personalization difficulty, defined by the number of environmental properties to be considered: the more properties involved, the higher the difficulty.

\newitem{Procedure.}
To bootstrap personalization, \sys{} requires device interaction logs to construct preference tables. However, since collecting such logs from each participant was practically infeasible within the limited experimental session, we randomly generated 100 interaction logs using GPT-4o. From these logs, \sys{} extracted preference tables over environmental properties and used them during the study. Prior to the experiment, participants were provided with an explanation of the preference tables used in the study and participated only after confirming their understanding. For comparison, \sage{} and \sasha{} were provided with the same interaction logs but handled them differently. \sage{} relied solely on them to infer device-specific preferences in line with its original design, while \sasha{}---originally lacking personalization---was extended by including this history in its prompt.

The experiment followed a within-subject design. To minimize order effects, the presentation order of the systems was randomized for each participant. First, in each session, all systems processed 8 tasks from scratch in \baseenv{}. Then, they re-executed the same tasks in \newenv{} with modified device configurations. After each phase, participants rated the outputs on three aspects using 7-point Likert scales (see Table~\ref{tab:preference_study_result}): perceived accuracy (Q1), device-level personalization (Q2), and parameter-level personalization (Q3). Q1 reflects participants' subjective perception of correctness, independent of ground truth, and thus complements the objective accuracy reported in Sec.~\ref{sec:performance_test}. Q2 and Q3 evaluate how effectively each system adapted outputs to user preferences at the device and parameter levels. System identities were blinded to prevent bias, and follow-up interviews captured participants' reasoning.

\subsubsection{Results}
Table~\ref{tab:preference_study_result} summarizes participant ratings for the three systems across \baseenv{} and \newenv{}. In Q1, \sys{} received the highest ratings in both environments, followed by \sage{} and \sasha{}. This trend is consistent with the objective accuracy results in Sec.~\ref{sec:performance_test}, suggesting that participants also perceived \sys{} as the most accurate system. For example, P2 noted that \sys{} ``\textit{clearly presented all the commands necessary for completing the task},'' while P11 criticized \sasha{} for producing ``\textit{many meaningless outputs with very low accuracy}.''

In Q2 and Q3, participants consistently rated \sys{} highest, emphasizing that it ``\textit{reflected user preferences as much as possible in both environments}'' (P4). By contrast, \sage{} was seen as only partially effective: P12 observed that it ``\textit{selected devices reasonably well, but the parameter values felt far off},'' and further noted that ``\textit{in the \newenv{} environment, it seemed to satisfy only some of the multiple preferences}.'' \sasha{} was rated the lowest, with P15 remarking that it ``\textit{ignored user preferences and instead made its own judgments about what services to provide}.'' These results highlight that \sys{} consistently delivered the strongest personalization, even under changing device configurations.

To examine statistical significance in the Q1, Q2, and Q3 scores across the three approaches, we conducted a Friedman test. In \baseenv{}, significant differences were observed for all three questions (Q1: $\chi^2(2)=24.03$, $p<0.001$, Q2: $\chi^2(2)=24.4$, $p<0.001$, Q3: $\chi^2(2)=24.4$, $p<0.001$). To investigate pairwise differences, we performed post-hoc Wilcoxon signed-rank tests with Holm–Bonferroni correction. The analysis showed significant differences across all pairs of approaches (all $p<0.05$). Similarly, in \newenv{}, significant differences were found for all three questions (Q1: $\chi^2(2)=24.4$, $p<0.001$, Q2: $\chi^2(2)=24.4$, $p<0.001$, Q3: $\chi^2(2)=24.4$, $p<0.001$). Post-hoc Wilcoxon signed-rank tests revealed significant differences across all pairs of approaches (all $p<0.05$), indicating that \sys{} substantially outperformed the two baseline systems in preference adaptation.

\subsection{Study 2: Overall Usability Evaluation}

We conducted a user study to evaluate the usability and interaction experience of \sys{} in realistic smart home scenarios. We compared user experiences with \sys{} against baseline approaches, focusing on whether our design choices translate into tangible benefits. Specifically, we sought to answer two research questions: \textit{i) Does the decomposition pipeline of \sys{} enable users to achieve more effective automation outcomes?} and \textit{ii) Is the hierarchical task memory of \sys{} effective in delivering reliable and fast responses?}

\subsubsection{Experimental Setup}
We set up our experimental environments as follows:

\newitem{Smart home settings.}
The smart home environments used in the study consisted of seven common IoT devices: a presence sensor, light, door lock, blind, air conditioner, fridge, and TV. 

\newitem{Approaches to be compared.}
We compared \sys{} with two baselines: the SmartThings app and \sage{}. We included the SmartThings app (tested on an Apple iPad 4) as a practical reference point for how users execute IoT commands today through a mainstream graphical interface. We chose \sage{} as the strongest research baseline among prior LLM-based agents, as it consistently achieved better accuracy and personalization than \sasha{} in our comparative evaluation. We evaluated \sys{} in two conditions to capture the effect of task memory on usability: \syscold{}, which starts with empty memory and represents a first-time user experience, and \syswarm{}, which is preloaded with task memory built from the 97 tasks in Sec.\ref{sec:performance_test} to represent continued use after memory accumulation. Both conditions used the same preference tables derived from the persona in Sec.\ref{sec:personalization_eval} because collecting interaction history per participant was impractical. Note that we used GPT-4o as the backbone model for all agents.

\newitem{Procedure.}
At the beginning of each experimental session, each participant was given an introduction to smart home technologies and conversational agents. We also explained the smart home configurations used in our study and how to use each approach, followed by a short tutorial that allowed participants to practice with the systems before the main tasks.

Participants were then asked to complete three task scenarios of increasing difficulty (\textit{easy}, \textit{medium}, and \textit{hard}) using all four approaches (SmartThings app, \sage{}, \syscold{}, and \syswarm{}). These scenarios were identical for all participants. For each scenario, we provided both a description and an example instruction (see Table~\ref{tab:user_study_scenario}), while allowing participants to phrase their instructions differently. All tasks were trigger-action rules and presented in order of increasing difficulty. To mitigate order effects across approaches, their sequence was counterbalanced across participants.

After each scenario, participants filled out a NASA-TLX questionnaire to assess perceived workload and were interviewed for immediate feedback. This process was repeated for all three scenarios across the four approaches. Finally, after completing all tasks, participants evaluated the usability of the four approaches using the System Usability Scale (SUS). The session concluded with an interview in which participants reflected on the strengths and weaknesses of each approach.

\begin{table*}[t]
    \centering
    \renewcommand{\arraystretch}{1.5} 
    \caption{Task Scenarios for the Usability Study, Grouped by Difficulty Level (Easy, Medium, Hard)}
    \label{tab:user_study_scenario}
    \begin{tabular}{c|m{10.5cm}|m{4.0cm}}
        \hline
        \textbf{Difficulty level} & \textbf{Scenario description} & \textbf{Example instruction} \\
        \hline
        \makecell{Easy \\ ($\geq 2$ devices, \\ explicitly specified)} & At 3 a.m., you wake up thirsty and quietly head to the kitchen. 
        You open the refrigerator, but the room is too dark to find a glass of water. 
        You think to yourself that it would be nice if the dining light automatically turned on whenever the refrigerator door was opened. 
        & ``\textit{Turn on the light in the dining room when I open the fridge}'' \\
        \hline
        \makecell{Medium \\ ($\geq 4$ devices, \\ unspecified)} & Every evening, you end your day by watching a movie on TV. To better focus on the experience, you usually set up the surrounding environment. However, repeating this preparation every time is starting to feel increasingly bothersome. 
        & ``\textit{When I watch TV, I want my house to be like a movie theater}'' \\
        \hline
        \makecell{Hard \\ ($\geq 6$ devices, \\ unspecified)} & After a long and tiring day, you head to your bedroom and set up the environment for sleep. However, as you repeat the same process every night, preparing it yourself at the end of the day is starting to feel increasingly tedious and bothersome. 
        & ``\textit{Make the room comfortable for me to sleep at night}'' \\
        \hline
    \end{tabular}
    \vspace{-0.3cm}
\end{table*}

\subsubsection{Results}
We now present the results of the second user study, which compared the four approaches across tasks of varying difficulty in terms of command generation time, perceived workload, and usability.

\begin{figure*}[t]
\centering
\subfloat[]{%
  \includegraphics[width=0.32\textwidth]{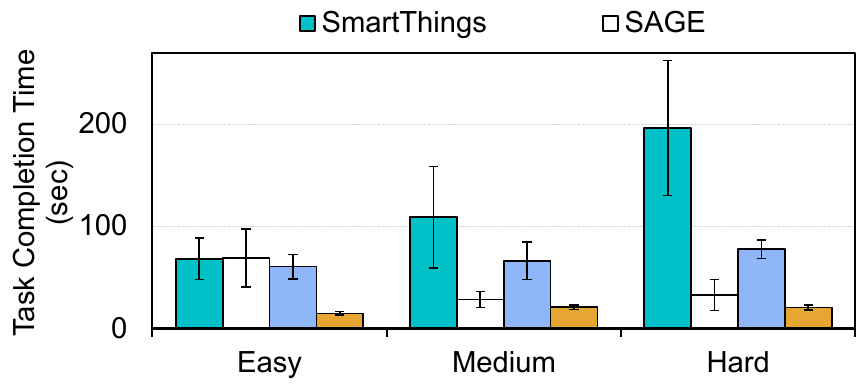}
}\hfil
\subfloat[]{%
  \includegraphics[width=0.32\textwidth]{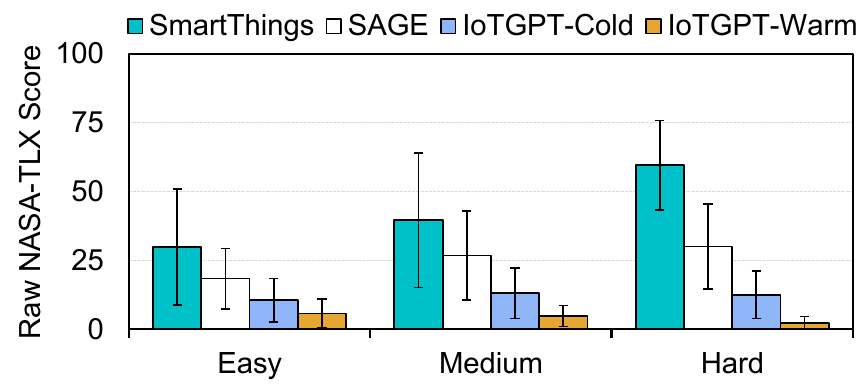}
}\hfil
\subfloat[]{%
  \includegraphics[width=0.33\textwidth]{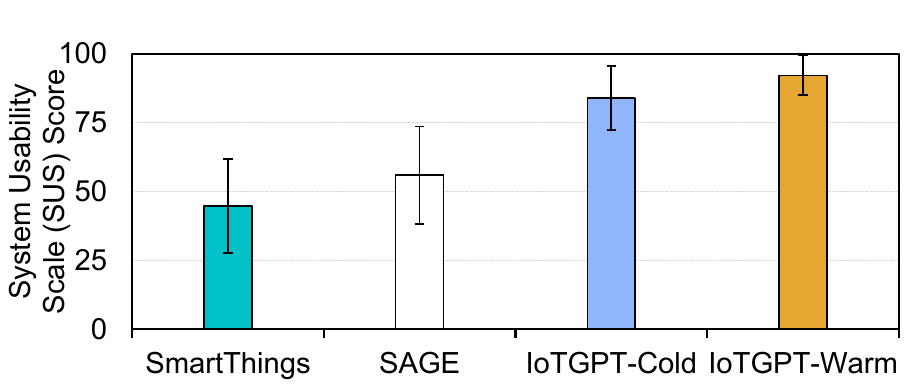}
}
\caption{(a) Average task completion time for each scenario. 
(b) Average raw NASA-TLX score for each scenario. 
(c) Average usability score of each approach. 
Each score was evaluated on a 100-point scale, and the error bars show the standard error.}
\vspace{-0.5cm}
\label{fig:us_average_score}
\end{figure*}

\newitem{Task Completion Time.}
We measured the time required to complete each automation task across the three scenarios. To account for different interaction modalities, we defined this metric differently for each approach: for the SmartThings app, it measures the human interaction time from initiating the automation setup to clicking the ``Save'' button, which applies the configured automation in the smart home environment. For the conversational agents (\sage{} and \sys{}), it represents the end-to-end latency from when a participant submits a natural-language instruction to when the generated commands are fully applied in the smart home environment.

As shown in Fig.~\ref{fig:us_average_score}(a), the SmartThings app consistently requires the longest times, except in the easy scenario. Moreover, its completion time increases sharply with task difficulty: in the hard scenario, the time is 187.7\% higher than in the easy scenario. This reflects the overhead of manually specifying every device action, which scales poorly as tasks grow more complex. Meanwhile, \sage{} shows the longest time in the easy scenario but exhibits an inverse trend with decreased times in the medium and hard scenarios. This counter-intuitive outcome stems from \sage{} misinterpreting trigger-action tasks as simpler direct-control commands, which reduced task completion time but led to inaccurate automations\footnote{Specifically, \sage{} failed to produce correct outputs for all participant tasks in the medium and hard scenarios, whereas both \sys{} variants consistently generated correct IoT commands across all scenarios.}. This underscores the risk of over-reliance on the LLM.

\syscold{} also exhibits increases in time as tasks grow more complex, but the rise is less steep than SmartThings (28.1\% increase from easy to hard). In contrast, \syswarm{} leverages task memory to achieve dramatically lower times across all scenarios, reducing time by 75.67\%, 68.54\%, and 73.46\% compared to \syscold{} in the easy, medium, and hard scenarios, respectively. These results highlight the efficiency gains from reusing previously learned subtasks.

We used a repeated-measures ANOVA to analyze differences in task completion time across approaches, applying Bonferroni correction in post-hoc tests to control for multiple comparisons. The ANOVA revealed a significant effect of approach on completion time for all scenarios (easy: $F(3,42)=31.84$, $p<0.001$; medium: $F(3,42)=32.59$, $p<0.001$; hard: $F(3,42)=79.19$, $p<0.001$).

In the easy scenario, post-hoc comparisons showed that \syswarm{} was significantly faster than the SmartThings app, \sage{}, and \syscold{} (all $p<0.001$). No other pairwise differences reached significance. For the medium and hard scenarios, we excluded \sage{} from statistical comparisons, as its misinterpretation of tasks led to invalid outputs. In the medium scenario, \syswarm{} outperformed both the SmartThings app and \syscold{} (all $p<0.001$), while no significant difference was found between the SmartThings app and \syscold{}. In the hard scenario, all pairwise comparisons were significant (all $p<0.05$). Across all scenarios, \syswarm{} consistently yielded the fastest time, and its advantage became more pronounced as task complexity increased.

\newitem{Perceived Task Workload.}
Fig.~\ref{fig:us_average_score}(b) shows the average raw NASA-TLX scores for each scenario. The SmartThings app consistently produced the highest workload, followed by \sage{}. \sage{}'s scores are largely due to poor task outcomes, as it often failed to produce satisfactory commands, leading to higher ratings in \textit{Performance} (indicating worse perceived outcomes) and \textit{Frustration}. In contrast, \syscold{} reduced workload substantially, and \syswarm{} further lowered it to the minimum across all scenarios. This suggests that \sys{} effectively minimizes cognitive effort, allowing users to manage complex tasks without feeling overwhelmed.

A Friedman test revealed a significant effect of approach across all scenarios (easy: $\chi^2(3)=18.69$; medium: $\chi^2(3)=26.49$; hard: $\chi^2(3)=37.54$; all $p<0.001$). Post-hoc Wilcoxon signed-rank comparisons with Holm–Bonferroni correction showed the following. In the easy scenario, \syswarm{} had significantly lower workload than the SmartThings app ($p=0.007$) and \sage{} ($p=0.009$), while the difference between \syscold{} and \syswarm{} was not significant ($p=0.129$). \syscold{} also yielded significantly lower workload than the SmartThings app ($p=0.017$) but not \sage{} ($p=0.0905$). In the medium scenario, \syswarm{} again had significantly lower workload than the SmartThings app ($p=0.005$), \sage{} ($p<0.001$), and \syscold{} ($p=0.016$). \syscold{} also showed significantly lower workload compared to the SmartThings app ($p=0.002$) and \sage{} ($p=0.025$). In the hard scenario, all pairwise comparisons were significant (all $p<0.05$).

\newitem{Usability.}
Fig.~\ref{fig:us_average_score}(c) shows the average SUS scores for each approach. The SmartThings app received the lowest ratings, followed by \sage{}, while both versions of \sys{} scored substantially higher. In particular, \syscold{} already reached the \textit{Excellent} range, and \syswarm{} pushed further into the \textit{Best Imaginable} category. These results highlight a clear advantage of \sys{} over both baselines in perceived usability.

A Friedman test revealed a significant effect of approach on usability ($\chi^2(3)=32.05$, $p<0.001$). Post-hoc Wilcoxon signed-rank comparisons confirmed that \syswarm{} was rated significantly more usable than the SmartThings app ($p<0.001$), \sage{} ($p=0.003$), and \syscold{} ($p=0.037$). \syscold{} also showed significantly higher usability than both the SmartThings app ($p=0.001$) and \sage{} ($p<0.001$).

\subsection{Overall Key Implications}
Our findings across both user studies yield four key implications for the design of smart home automation systems.

\newitem{Manual vs. Conversational Interfaces.}
Participants consistently found the SmartThings app burdensome as task complexity increased. For example, P3 noted, ``\textit{The existing platform feels too complicated and difficult because there are too many items to select}.'' Similarly, P13 commented, ``\textit{I think this would be inaccessible for older users}.'' When faced with more complex tasks, participants struggled with confidence and clarity; P6 admitted, ``\textit{I wasn't sure if I was setting things correctly},'' and P5 stated, ``\textit{I didn't know what to do to make the house feel like a movie theater}.'' Taken together, these remarks highlight the scalability limits of manual, visual interfaces.

In contrast, conversational agents were perceived as more natural and efficient, especially for complex tasks. Participants praised \sys{}: P1 observed, ``\textit{Controlling multiple devices with just one spoken command is very intuitive and convenient}.'' P7 added, ``\textit{It's impressive that the system suggests device controls I wouldn't have thought of myself}.'' Another participant emphasized how \sys{} could handle vague commands: P9 reflected, ``\textit{It's nice that it can naturally concretize and execute something as vague as `make it like a movie theater.' Even I would have to think hard about what exactly that means for me}.'' These findings underscore the value of natural-language interaction for smart home automation.

\newitem{Risks of Over-Reliance on LLMs.}
While existing LLM-based agents supported natural interactions, their heavy reliance on the LLM led to unstable outcomes and slower responses. \sage{}, in particular, replans reasoning modules from scratch for each task and queries the LLM iteratively until completion, making it highly dependent on the LLM. This not only prolongs execution but also increases the likelihood of errors when tasks become more complex or ambiguous. 
Participants highlighted these drawbacks directly. P2 remarked, ``\textit{Even for simple tasks, it takes longer than I expected}.'' P10 noted, ``\textit{It ignores the condition (trigger) in my command and only generates the action part, which makes it impractical in real life}.'' P8 shared a more serious concern: ``\textit{The system suddenly raised the TV volume to 100. There needs to be a safeguard for this kind of behavior}.'' Together, these findings highlight the need for complementary mechanisms---such as task memory---to reduce repeated reliance on the LLM and ensure more reliable performance.

\newitem{Cold-Start vs. Warm-Start Usability.}
Our findings show that \sys{} provides strong usability even in the cold-start condition, where no prior task memory exists. Participants were still able to complete complex automation tasks with relatively low workload and high satisfaction, suggesting that the system's decomposition pipeline already delivers meaningful benefits without prior learning. However, the warm-start variant further amplified these strengths: by reusing accumulated task memory, it reduced command generation time, minimized workload, and improved overall usability. This demonstrates that while \sys{} is immediately usable out of the box, its usability improves incrementally as task memory grows, offering a clear path for long-term adoption.
Participants' feedback reinforced these observations. P9 noted, ``\textit{Even without prior learning, it still worked well and felt usable}.'' P12 emphasized the incremental improvement, ``\textit{The earlier version (\textup{\syscold{}}) took longer than I expected, but the next version (\textup{\syswarm{}}) delivered the desired results quickly, which was very satisfying}.'' Similarly, P10 added, ``\textit{The earlier version worked well enough to be useful, but the learned version was fast enough to handle immediate commands, making it feel far more practical}.''

\newitem{The Importance of Personalization.}
Our studies highlight personalization as a core requirement for smart home automation. Participants valued systems that could tailor commands to their preferences and contexts, noting that such support made automation feel more practical and trustworthy. For example, P10 explained, ``\textit{It was convenient that multiple devices were adjusted to the given preferences without me having to configure them manually}.'' Similarly, P7 remarked, ``\textit{It seemed that the system could even suggest useful device usage patterns that I hadn't recognized myself.}.''
Beyond static personalization, the results of Study 1 point to the importance of adaptiveness: when device configurations changed, systems that mapped preferences at the level of environmental properties preserved user preferences more effectively. As P4 commented, ``\textit{It was impressive that the system could still reflect my preferences even when the devices were different}.'' Together, these findings underscore that personalization---particularly adaptive personalization---is essential for sustaining usability in dynamic home environments.

\section{Limitation and Future Work}

\newitem{Dilemma of Conflicting User Preferences.}
Personalization can become ambiguous when user preferences across different environmental properties conflict. For example, a user who prefers both low temperature and low noise may face a dilemma: running the air conditioner cools the room but introduces noise as a side effect. In such cases, it is unclear whether the system should prioritize temperature or noise. Since our current prototype, built on the EUPont ontology, does not account for such side effects, it defaults to turning on the air conditioner. However, this can result in suboptimal personalization if the user values quietness over temperature. To address this, future work should extend the ontology to represent device side effects and refine preference granularity, enabling the system to resolve such conflicts more effectively.

\newitem{Need for Local Inference.}
While \sys{} reduced command generation latency through task memory, some participants still perceived delays. For instance, P4 remarked, ``\textit{Although the execution time was optimized, it still feels long}.'' Beyond performance, concerns about data privacy also emerged. As P6 noted, ``\textit{Natural-language commands may unintentionally include private information, which could be exposed if processed by a cloud-based LLM}.'' These observations point to the need for future work on enabling local inference. Developing lightweight LLMs tailored for smart home automation could both reduce latency and mitigate privacy risks, making such systems more practical and trustworthy in everyday use.

\newitem{Limitations in Context Awareness.}
A limitation of \sys{} is that it infers context only from user instructions, which cannot fully capture situational factors. For instance, combining activity context (e.g., recognizing whether the user is resting or exercising through home camera images analyzed by a VLM) with environmental state context (e.g., temperature, humidity, or noise measured by sensors) could enable more accurate refinement of IoT commands. However, incorporating such multimodal context risks over-fragmenting task memory, making reuse of personalization knowledge difficult. Future work should therefore explore abstraction techniques---such as clustering or hierarchical representation of context---to capture diverse situations while maintaining reusability.

\newitem{From Manual Checking to Built-in Safety.}
In our current prototype, the responsibility for filtering unsafe or hallucinated commands lies with the user, who must review the generated automation and provide corrective feedback if necessary. While this approach prevents immediate risks, it imposes additional burden on the user and reduces usability. To address this limitation, future work should explore automated safety mechanisms that proactively verify or adjust commands before execution. For example, if the system mistakenly generates a command to raise the TV volume to 100, a built-in safeguard could automatically limit the value to a safe threshold (e.g., 50) without requiring user intervention. Such verification and fail-safe measures would improve reliability while reducing user burden, allowing \sys{} to deliver safe automation with minimal oversight.

\newitem{Long-term User Experience.}
Our evaluation was limited to short-term usage, as participants interacted with \sys{} for about 60 minutes during the user study. While these sessions provided useful insights into usability and personalization, they cannot fully capture how users might adapt to or rely on \sys{} in daily life over longer periods. Future work should conduct longitudinal studies to examine how sustained use affects user trust, system adaptability, and overall satisfaction in real-world smart home environments.

\section{Conclusion}
We introduced \sys{}, an LLM-based smart home agent that leverages task decomposition and task memory to generate IoT commands in a reliable, efficient, and personalized manner. \sys{} combines structured reasoning with memory reuse to reduce unreliability, latency and cost in IoT command generation, while also supporting adaptive personalization through device-agnostic preference modeling. Our evaluation across performance benchmarks and user studies demonstrates improvements in accuracy, efficiency, and usability compared to baselines. We believe that \sys{} represents a concrete step toward more dependable and user-centered AI agents, with implications extending beyond smart homes to broader contexts where reliable and adaptive automation is essential.

\bibliographystyle{IEEEtran}
\bibliography{references}

\end{document}